\documentclass[a4paper,fleqn]{cas-dc}
\usepackage[numbers]{natbib}

\def\tsc#1{\csdef{#1}{\textsc{\lowercase{#1}}\xspace}}
\tsc{WGM}
\tsc{QE}
\usepackage{hyperref}
\usepackage{xspace}
\usepackage{booktabs}
\usepackage{graphicx}
\usepackage{multirow}
\usepackage{enumitem}
\usepackage{url}
\usepackage{amsmath}
\usepackage{amssymb}

\usepackage[ruled,linesnumbered]{algorithm2e}

\newcommand{\etal}{\textit{et al.}\xspace}
\newcommand{\blue}[1]{{\color{black}#1}}

\begin{document}
\let\WriteBookmarks\relax
\def\floatpagepagefraction{1}
\def\textpagefraction{.001}
\let\printorcid\relax 

\shorttitle{}    

\shortauthors{Ruitong Liu et al.}

\title[mode = title]{Vul-LMGNNs: Fusing Language Models and Online-Distilled Graph Neural Networks for Code Vulnerability Detection}  

\tnotetext[1]{This work was supported in part by the National Natural Science Foundation of China under Grant 61902342, and in part by the Defense Industrial Technology Development Program under Grant JCKY 2021602B002.}

\author[1,4]{Ruitong Liu}
\cormark[0]
\ead{liuruitong@bupt.edu.cn}

\credit{Conceptualization of this study, Methodology, Software}

\affiliation[1]{organization={School of Cyber Science and Technology, College of Computer Sciece and Technology, Zhejiang University},
                city={Hangzhou},
                postcode={330000}, 
                country={China}}

\author[1]{Yanbin Wang }[type=editor,auid=000,bioid=1,orcid=0000-0003-1682-5712]
\cormark[1]
\ead{wyb@zuaa.zju.edu.cn}

\credit{Data curation, Writing - Original draft preparation}

\affiliation[2]{organization={Hangzhou Research Institute of Xidian University},
                postcode={310027}, 
                city={Hangzhou},
                country={China}}

\author[1]{Haitao Xu }
\cormark[1]
\ead{haitaoxu@zju.edu.cn}

\author[2]{Jianguo Sun }
\cormark[0]
\ead{jgsun@xidian.edu.cn}

\author[1]{Fan Zhang }
\cormark[0]
\ead{fanzhang@zju.edu.cn}

\author[3]{Peiyue Li }
\cormark[0]
\ead{lipeiyue@ppsuc.edu.cn}

\author[1]{Zhenhao Guo }
\cormark[0]
\ead{guozhenhao17@mails.ucas.ac.cn}


\affiliation[3]{organization={People's Public Security University of China},
                city={Beijing},
                postcode={100038}, 
                country={China}}

\affiliation[4]{organization={Shenzhen MSU-BIT University},
                city={Shenzhen},
                postcode={518172}, 
                country={China}}
                
\cortext[cor1]{Corresponding author}

\begin{abstract}
Code Language Models (codeLMs) and Graph Neural Networks (GNNs) are widely used in code vulnerability detection. However, a critical yet often overlooked issue is that GNNs primarily rely on aggregating information from adjacent nodes, limiting structural information transfer to single-layer updates. In code graphs, nodes and relationships typically require cross-layer information propagation to fully capture complex program logic and potential vulnerability patterns. \blue{Furthermore, while some studies utilize codeLMs to supplement GNNs with code semantic information, existing integration methods have not fully explored the potential of their collaborative effects.}

To address these challenges, we introduce Vul-LMGNNs that integrates pre-trained CodeLMs with GNNs, leveraging knowledge distillation to facilitate cross-layer propagation of both code semantic knowledge and structural information. Specifically, Vul-LMGNNs utilizes Code Property Graphs (CPGs) to incorporate code syntax, control flow, and data dependencies, while employing gated GNNs to extract structural information in the CPG. \blue{To achieve cross-layer information transmission, we implement an online knowledge distillation (KD) program that enables a single student GNN to acquire structural information extracted from a simultaneously trained counterpart through an alternating training procedure.} \blue{Additionally, we leverage pre-trained CodeLMs to extract semantic features from code sequences.} Finally, we propose an "implicit-explicit" joint training framework to better leverage the strengths of both CodeLMs and GNNs. In the implicit phase, we utilize CodeLMs to initialize the node embeddings of each student GNN. Through online knowledge distillation, we facilitate the propagation of both code semantics and structural information across layers. In the explicit phase, we perform linear interpolation between the CodeLM and the distilled GNN to learn a late fusion model. The proposed method, evaluated across four real-world vulnerability datasets, demonstrated superior performance compared to 17 state-of-the-art approaches. Our source code can be accessed via GitHub: \url{https://github.com/Vul-LMGNN/vul-LMGGNN}.

\end{abstract}




\maketitle

\section{Introduction}
With the rapid expansion of the open-source community, software vulnerability detection technology has become a significant concern in the software industry and cybersecurity domain. Vulnerabilities pose a threat to the integrity and availability of software and computer systems, potentially leading to privilege escalation, leakage of sensitive data, denial of service, and various other attacks, resulting in substantial economic and societal losses \cite{plate2015impact,lin2020software}. \blue{Traditionally, developers and security engineers rely on rule-based code analysis and symbolic execution techniques \cite{lipp2022empirical}. Although these methods can detect vulnerabilities effectively, they often generate high false-positive rates, requiring extensive manual verification.} 

To improve the efficiency of code vulnerability detection, extensive research has leveraged deep learning (DL) models for automated vulnerability detection. These methods extract features from the source code to generate initial embedding vectors, which are then fed into neural networks to learn vulnerability patterns and produce classification results, thereby achieving automatic detection capabilities \cite{russell2018automated}.

Deep learning for code vulnerability detection primarily falls into two categories: sequence-based and graph-based approaches. Sequence-based methods convert source code or its structures into serialized forms, treating elements as tokens \cite{wu2022code,10.1145/3597926.3598037}. These approaches employ various architectures such as LSTMs \cite{lin2019software,li2018vuldeepecker}, CNNs \cite{liang2019jsac}, or modern language models to detect vulnerabilities by learning sequential features. Graph-based methods, conversely, transform source code into heterogeneous graph structures like Abstract Syntax Trees (AST), Control Flow Graphs (CFG), and Program Dependence Graphs (PDG). These methods then employ Graph Neural Networks (GNNs) to capture both local structures and dependencies, offering richer syntactic and structural insights compared to code sequences \cite{wang2020combining,lin2019software}.

However, current work has two main limitations: First, current GNN-based methods for vulnerability detection suffer from a common weakness: the GNNs they employ are inherently limited in their ability to propagate structural knowledge across multiple layers of the graph. This limitation stems from the fundamental architecture of typical GNNs, which primarily aggregate information within a single layer during each iteration of the message-passing process. Even in multi-layer GNNs, where multiple rounds of aggregation are performed, each layer's updates are still based on immediate neighbors, failing to directly capture long-range dependencies or complex structural patterns that span across multiple layers of the graph. This localized information processing can be particularly problematic in the context of code analysis, where vulnerabilities often arise from intricate interactions between distant parts of the code structure, requiring a more holistic understanding of the program's architecture and data flow. Secondly, GNNs offer distinct strengths in code analysis. Sequence-based approaches excel at capturing contextual semantics, while GNNs provide valuable structural insights. However, current research rarely leverages the potential synergy between these complementary approaches.

To address these challenges, we propose Vul-LMGNNs, a joint training framework that effectively integrates codeLMs and GNNs, while resolving information transfer limitations in code graph learning. The core of Vul-LMGNNs is a two-stage fusion scheme implementing an "implicit-explicit" integration: 1) Implicit Stage: We utilize semantic embeddings extracted by codeLMs to initialize nodes in the code GNN. This approach fuses semantic information with code structural information during graph update learning, constituting the initial information integration. 2) Explicit Stage: We employ linear interpolation to combine the predictions of both model types, achieving the final model fusion. Furthermore, by integrating online knowledge distillation into the first stage of our fusion architecture, we simultaneously facilitate the effective cross-layer information propagation of both semantic and graph structural information.

The contributions of our work can be summarized as follows:

\blue{
\begin{itemize}[left=2pt,label=$\bullet$,itemsep=2pt]

\item We propose a novel approach for detecting code vulnerabilities that leverages codeLMs to learn the semantics from source code sequences and employs gated GNN to grasp syntactic, control, and data dependencies from code attribute graphs.

\item We propose an "implicit-explicit" dual-learning framework for the systematic integration of CodeLMs and GNNs. In the implicit stage, we perform an embedding-level fusion of the two models. In the explicit stage, we employ linear interpolation for the output consolidation at the model endpoints.

\item We propose the incorporation of an online knowledge distillation module into the joint training framework of CodeLMs and GNNs to address the limitations of cross-layer knowledge propagation in code graphs and to further facilitate the integration of code semantic information with code structural information.

\item The proposed method is extensively evaluated on four public datasets, achieving state-of-the-art performance, with approximately a 10\% improvement in F1 scores on two challenging imbalanced datasets compared to the previous most competitive baseline methods.

\end{itemize}
}

The rest of this paper is structured as follows: Section \ref{sec:related} provides an overview of the background and related work. Section \ref{sec:dataset} outlines the composition of the dataset. Section \ref{sec:framework} delves into the design specifics of our model. Section \ref{sec:results} presents the experimental outcomes and evaluates our model’s performance relative to the baseline method across datasets. Finally, Section \ref{sec:conculsion} summarizes the paper and outlines directions for future research.

\section{Related Work}
\label{sec:related}
In this section, we review the most relevant works to our study, focusing on those based on deep learning techniques. These can be broadly categorized into two groups: sequence-based approaches and graph-based approaches.

\subsection{Sequence Based Models}
Current studies based on deep sequence models generally follow the process of preprocessing, vectorization, and neural network modeling \cite{wu2022code,russell2018automated,harer2018automated,8846081,9416836}. In data preprocessing, the raw source code is subjected to slicing and normalization techniques, after which it is parsed into a sequence of tokens. Subsequently, these tokens are transformed into vectors suitable for neural network processing. RNN and transformer-based models are used to learn contextual information within token sequences and to make the final defect prediction. RNN-based works, such as VulDeePecker \cite{li2018vuldeepecker} and SySeVR \cite{li2021sysevr}, have introduced lexical analysis, which converts the source code into a more fine-grained code snippet. A potential concern with code slicing is that the extracted code representations may not encompass all vulnerable code snippets. On the contrary, transformer-based methods utilize token vectorization techniques that extract more vulnerability-aware features. Transformer-based approaches often omit code slicing and normalization strategies, opting instead to directly tokenize the source code. Chen \etal evaluated the performance of several Transformer-based methods, including CodeBERT\cite{feng2020codebert} and GraphCodeBERT\cite{guo2020graphcodebert}, on the DiverseVul dataset for vulnerability detection. Other methods have adopted different tokenization strategies from the NLP domain; for instance, CodeT5 \cite{wang2021codet5} uses byte-level byte-pair-encoding (BPE) \cite{radford2019language} to segment the code into tokens, while CoTEXT \cite{phan2021cotext} opts for the Sentencepiece \cite{kudo2018sentencepiece} model to extract tokens. These methods have been proven to be effective.

\begin{figure*}
    \centering
    \includegraphics[width=\linewidth]{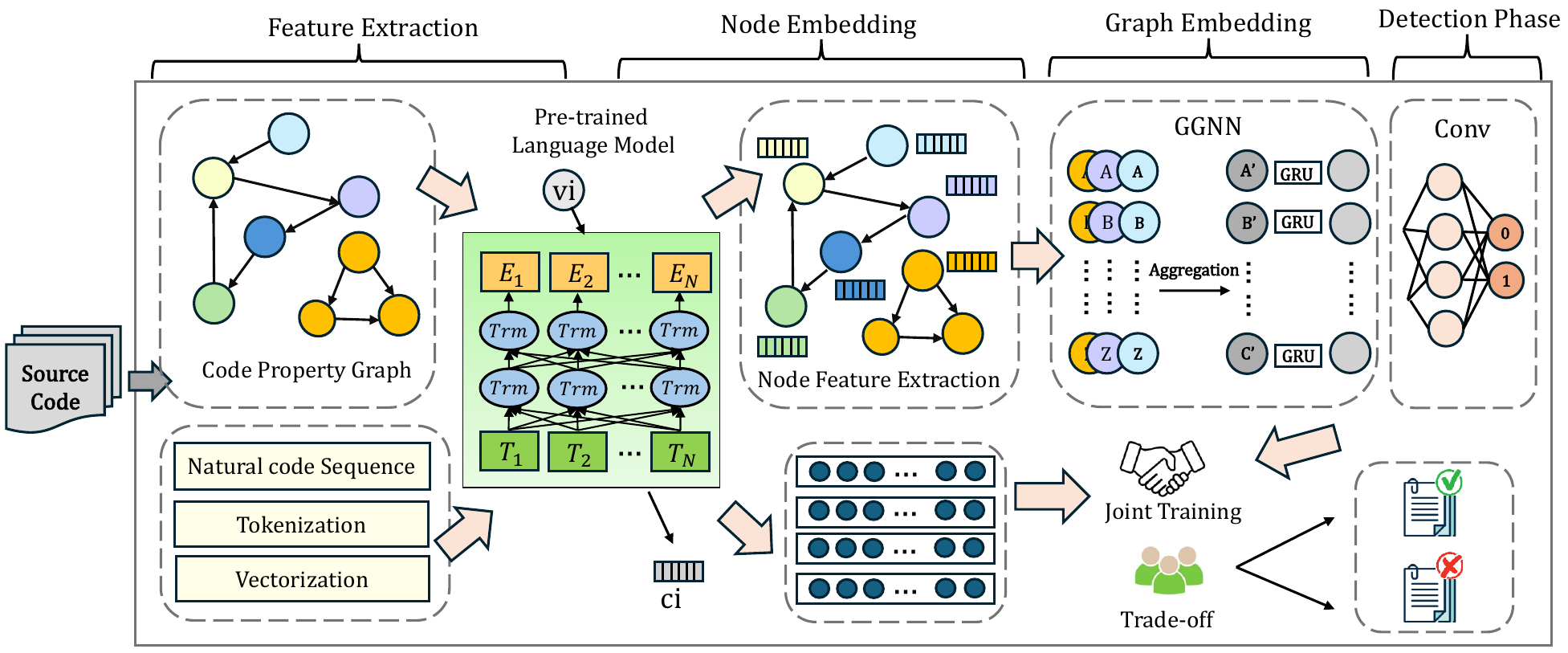}
    \caption{Overview of the Vul-LMGNNs Vulnerability Detection Framework.}
    \label{fig:overview}
\end{figure*}

\subsection{Graph Based Models}
In the field of code vulnerability identification, GNN-based approaches leverage various code representations as input data, including AST, CFG, PDG, and CPG\cite{yamaguchi2014modeling,wu2022code}, and then employ graph embedding techniques or graph representation learning methods\cite{yuan2023continual}, such as Graph Convolutional Networks (GCNs), Graph Attention Networks (GATs), knowledge Graph \cite{ni2024knowledge}. These techniques aim to transform the complex, high-dimensional graph structures into compact, low-dimensional vector representations while preserving essential structural and semantic information \cite{nguyen2022regvd,cheng2022path,10.1145/3597926.3598145,li2021vulnerability,hin2022linevd}. Methods like AI4VA \cite{suneja2020learning} and those proposed by Feng \etal \cite{feng2020graph} directly use the original versions of the four basic graphs as their code representations. The Devign \cite{zhou2019devign} was the first to employ a GNN for code vulnerability detection tasks, incorporating Natural Code Sequence (NCS) edges into the CPG. Chakraborty \etal \cite{chakraborty2021deep} proposed the ReVeal algorithm, which combines gated GNNs with multilayer perceptrons; FUNDED \cite{wang2020combining} introduced an enhanced AST with eight additional edge types. Unlike the aforementioned strategies that add structural information, VulSPG \cite{zheng2021vu1spg} suggests eliminating code unrelated to vulnerabilities. It performs graph slicing on the CPG to generate the SPG. 

\blue{Sequence-based models excel at learning code semantics but struggle to capture the non-linear, hierarchical structures in code that involve control flow, nested functions, and variable dependencies. GNNs excel at learning from graph-structured data but are often constrained by single-layer structural information transfer. Moreover, when applied to code graphs, they struggle to incorporate the contextual semantic information of the original code. This insight motivated our proposal to combine pre-trained code language models, code graph models, and online knowledge distillation.
}

\begin{figure}
    \centering
    \includegraphics[width=0.85\linewidth]{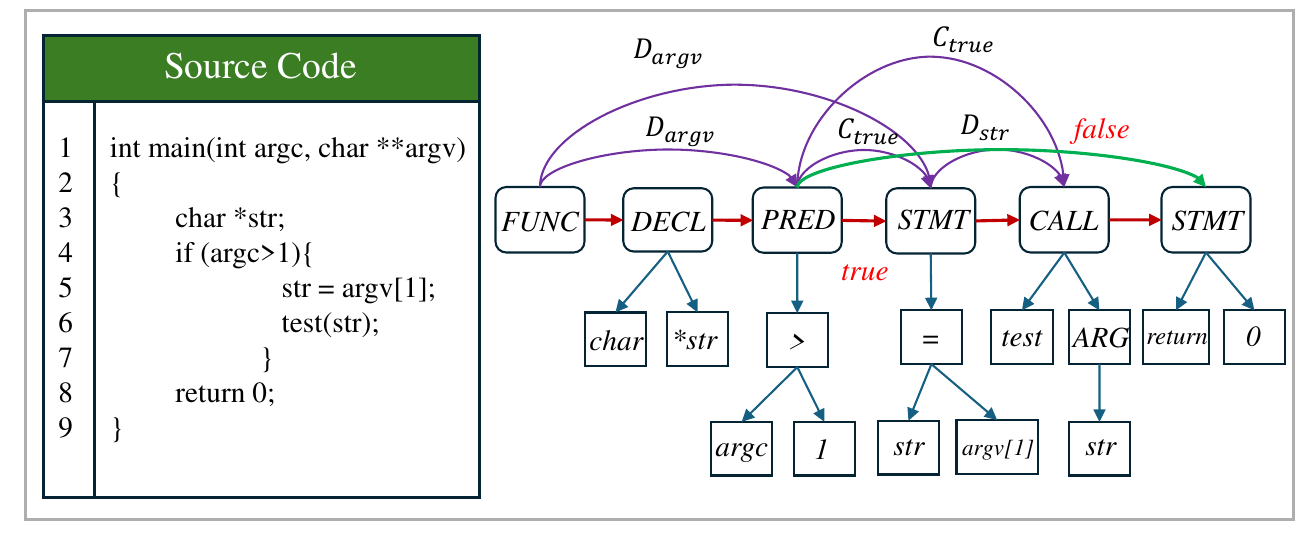}
    \caption{\textbf{A CPG for the example source code. Edge-type legend:\textit{ Blue = AST, Red = CFG, Purple = PDG}.}}
    \label{cpg}
\end{figure}

\section{Vul-LMGNNs}
\label{sec:framework}
In this section, we provide a detailed explanation of the Vul-LMGNNs process and its components. For clarity, our explanation is divided into several parts:

\begin{enumerate}
    \item \textbf{Code Representation:} We utilize two perspectives for code representation:
    \begin{itemize}
        \item Code graph representation
        \item Code sequence representation
    \end{itemize}
    We describe how to process these to produce the corresponding representations.

    \item \textbf{Code Property Graph Creation:} We employ code property graphs to transform source code into an abstract graph representation. We introduce this code property graph here.

    \item \textbf{Gated GNN:} For learning on the code property graph, we adopt a gated graph neural network. Its advantage lies in dynamically controlling information transmission and retention.

    \item \textbf{Online Knowledge Distillation for Graph Neural Networks:} This part describes how we leverage online knowledge distillation to enable cross-layer knowledge transfer in the gated GNN, thereby improving its performance.

    \item \textbf{Code Sequence Semantic Information Extraction:} Here, we use pre-trained code language models to extract contextual semantic embeddings from code sequences.

    \item \textbf{Joint Training:} This section outlines how we jointly train the code language model and GNN to achieve implicit and explicit information fusion. This part will integrate the modules from sections 3, 4, and 5.
\end{enumerate}

\begin{algorithm}
\caption{Vul-LMGNNs: Code Vulnerability Detection}

\KwIn{Train data - \( D_{\text{train}} \)}
\Indp \Indp  
Contribution of triplet loss - \( \alpha \)\\
Contribution of regularization loss - \( \beta \)\\
Separation boundary - \( \gamma \)\\
Learning rate - \( lr \)\\
Tradeoff parameter - \( \lambda \)\\
\Indm \Indm
\KwOut{Trained model}

\SetKwFunction{FMain}{Vul-LMGNNs}
\SetKwProg{Fn}{Function}{:}{}
\Fn{\FMain{}}{
    \( \text{features} \gets \emptyset \)\\
    \( \text{labels} \gets \emptyset \)\\
    \(\triangleright\) Features extraction process \\

    \For{\( (C,L) \in D_{\text{train}} \)}{
    \( (V, E) \gets \text{extract\_code\_property\_graph}(C) \)\\
    \For{\( v \in V \)}{
        \( T_v \gets \text{onehot}(v.\text{type}()) \)\\
        \( C_v, S_v \gets \textit{CodeLM}(v.\text{fragment}(), C) \)\\
        \( x_v \blue{\gets} \text{concat}(T_v, C_v) \)\\
    }
     \( \tilde{X} \blue{\gets} \textit{Online-Distilled GGNN}(x_v,E) \)\\
     \( x_g \blue{\gets} \textit{Aggregate}(\tilde{X}) \)\\
        \( \text{features} \gets \text{features} \cup x_g \cup S_v\)\\
        \( \text{labels} \gets \text{labels} \cup L \)\\
    }

   \( M \gets \textit{Combined-RepresentationModel}() \)\\
    \(\triangleright\) Model training process\\
    \For{\( (f_g, l_g) \in D_{\text{train}} \)}{
        \(\triangleright\) Define the loss function.\\
        \( L_{\text{all}}\gets\text{loss\_function}(M, D_{\text{train}} , f_g , l_g , \alpha, \beta , \gamma, \lambda) \)\\
        \( \theta \) represents the model parameters of \( M_{Combined} \).\\
        \( \theta \gets \theta - \nabla_{\theta}(L_{\text{all}}) \)\\
    }
    \KwRet \( M_{\theta} \)
}
\end{algorithm}

\subsection{Code Representation}
The purpose of this phase to transform the original function-level source code into fixed-length feature vectors that contains both semantic and syntactic structural information. Such conversion prepares the suitable data format for efficient processing by GNN models and code language models that follow. To achieve this, we adopt two specialized code representation strategies for GNNs and code sequence language models.

\textbf{For code graph representation:} We employ the open-source code analysis tool Joern \cite{joern} to parse the source code and generate the CPG. This CPG provides a unified and concise representation that combines control and data flow with abstract syntax trees and dependency graphs. We rigorously exclude functions with errors in graph generation to ensure data quality.

\textbf{For code sequence representation:} we adhere to the approach presented in \cite{zhou2019devign}, by converting function-level code into natural code sequences. This method serializes the code in alignment with the natural order of the source code, thereby preserving the logical sequence of the code.

\subsection{Code Property Graphs}
In the process of generating CPG, functions are transformed into comprehensive graphs that comprise various types of nodes, such as variables and function calls, and edges, including control flow and data flow, which convey distinct types of information. At the core of the CPG, the AST captures the syntactic information, modeling the hierarchical structure of functions in a way that outlines the grammar and composition of the code. However, since the AST primarily offers a static representation, it lacks the capacity to infer the program’s dynamic behavior. To address this, CPGs incorporate additional types of edges to represent data flow and control flow, thereby enriching the graph with insights into the execution context and dependencies between code segments. This integration frames a more holistic understanding of both the static structure and dynamic behavior of the program.

The representation of the CPG is denoted as $G = (V, E)$, where $V$ represents the nodes within the graph and $E$ means the edges. Each vertex $V$ in the CPG encompasses the vertex type and a segment of the original code. As illustrated in the Fig. \ref{cpg}, the nodes and blue edges represent the AST structure of this function segment, with the purple edges marked ``$D_{argv}$'' indicating the data dependency from the subtree defining variable $argv$ to the subtree using the defined value. The red edges denote the execution order within the function.

For the node set $V$, every node $v \in V$ can contain various types of information depending on its source, such as AST, CFG, or PDG. This includes CPG node type identifiers such as $IdentifierDeclType$ or keywords such as $int, char, for$, or operators such as $+, -$.

\begin{figure}[t]
    \centering
    \includegraphics[width=1.0\linewidth]{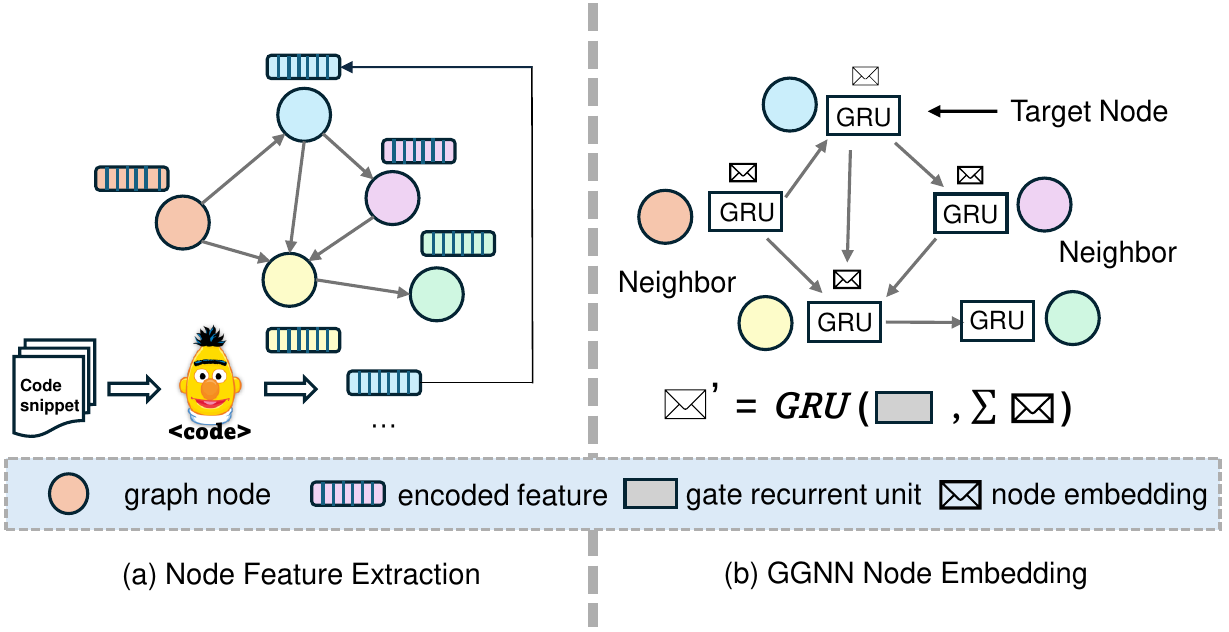} 
    \caption{\textbf{Feature extraction and node embedding phases.}}
    \label{feature}
\end{figure}

\begin{figure}[t]
    \centering
    \includegraphics[width=1.0\linewidth]{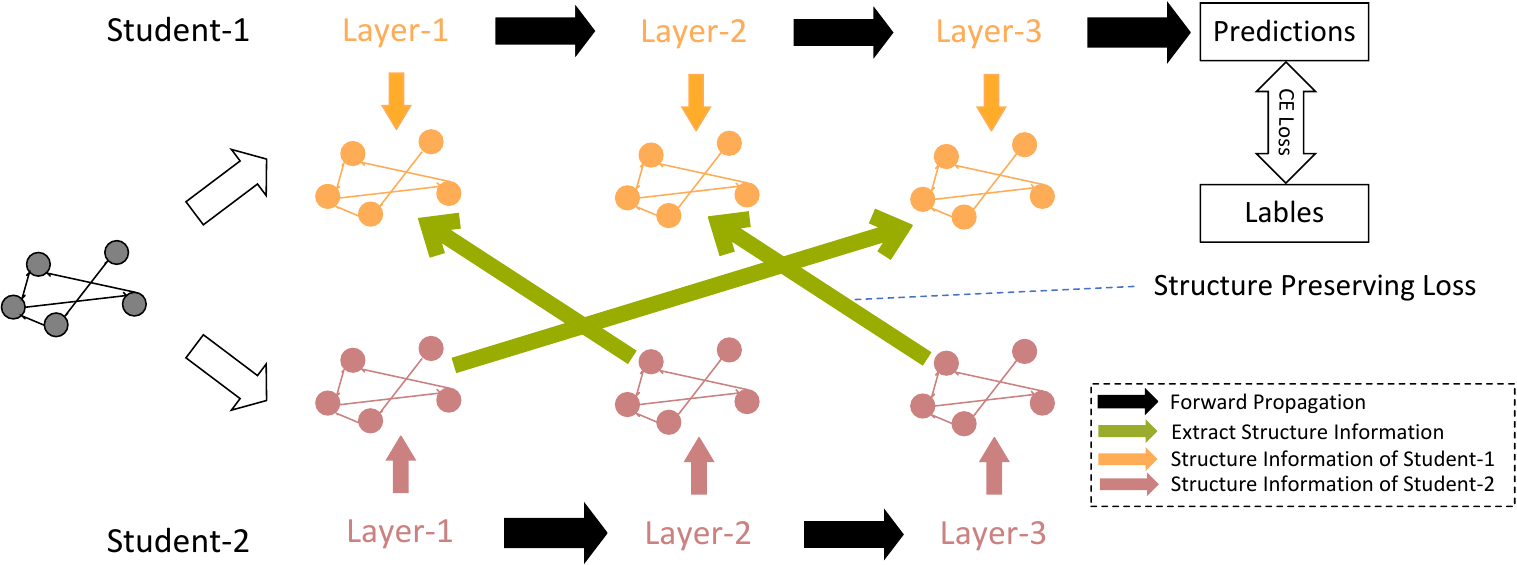} 
    \caption{\textbf{Schematic diagram of online knowledge distillation based on knowledge alignment.}}
    \label{KD}
\end{figure}

\subsection{Gated Graph Neural Network}
In this section, we leverage GGNNs to explore CPGs, utilizing their advanced capabilities to discern patterns of information flow across nodes, thereby revealing structural insights pertinent to code properties.

GGNNs are fed with feature vectors of all the nodes alongside the graph edges. For a specified embedded graph $g_i(V, X, A)$, where $V$ indicates nodes, $X$ their features, and $A$ the adjacency relationships,the GGNN assigns a Gated Recurrent Unit (GRU) to each node $v_j \in V$. This GRU updates the current vertex embedding by integrating the embeddings of all its neighboring nodes. Specifically, the initial state vector for a node $h^{(1)}_j \in \mathbb{R}^z$, where $z \geq d$, is initialized by copying $x_j$ into the first dimensions and padding with additional zeros. To update node embeddings, we employ a neighborhood aggregation scheme. At each node, messages are aggregated and subsequently utilized to update the associated node representation at the subsequent embedding layer. Formally,
\begin{equation}
a^{t}_{v,g} = A^{T}_{(v,g)} \left( \left[ h^{(t-1)T}_1, \ldots, h^{(t-1)T}_m \right] + b \right)
\end{equation}
To be specific, $t$ represents a specific time step, $b$ denotes the bias vector, and $A$ is the adjacency matrix. The subsequent state $a^{t}_{v,g}$ of node $v_j$ is computed by aggregating the information from all neighboring nodes as defined in the adjacency matrix $A_(v,g)$ for a particular edge type.

Subsequently, the GRU algorithm is used to aggregate and update the states for identical nodes across different graphs. The process is articulated as follows:
\begin{equation}
 z^{t}_{v,g} = \sigma(W^{z} \cdot AGG(a_{v,g}^t) + U^{z}h_{v,g}^{(t-1)}) 
\end{equation}

\begin{equation}
 r^{t}_{v,g} = \sigma(W^{r} \cdot AGG(a_{v,g}^t) + U^{r}h_{v,g}^{(t-1)}) 
\end{equation}

\begin{equation}
\widetilde{h^t_{v,g}} = \text{tanh}(W \cdot AGG({a}_{v,g}^t) + U(r^t_{v,g}\circ h^{(t-1)}_{v,g}))
\end{equation}

\begin{equation}
 h^t_{v, g}= (1-z^t_{v, g})\circ h^{(t-1)}_{v, g}+ z^t_{v, g}\circ \widetilde{{h}^t_{ v, g}} 
 \end{equation}
 Where $h^{(t-1)}_{v,g}$ is the hidden state of node $v$ in graph $g$, $z^{t}_{v,g}$ and $r^{t}_{v,g}$ are the update gate and reset gate, respectively. $\hat{h}^{t}_{v,g}$ is the candidate hidden state, and $h^{t}_{v,g}$ is the output hidden state. $AGG$ denotes the aggregation function, which is utilized to compile information from various edge types. In our application, we have employed the SUM \cite{zhou2019devign} function.
 
 The final step involves aggregating all vertex embeddings into a single vector to represent the entire CPG. Specifically,
 
\begin{equation}
H^{(T)}_{(v,g)} = \sum_{v \in V} h^t_{v, g}
\end{equation}

 Subsequently, we adopt a training mechanism similar to that of \cite{zhou2019devign,russell2018automated}, which deconstructs the task into 'learning code representation' and 'learning vulnerability'. This approach introduced an output layer designed to highlight the nodes with the most significant information for the task of vulnerability detection. We utilized convolution and max-pooling operations, commonly employed in CNNs. $\alpha(\cdot)$ is defined as a one-dimensional convolutional layer accompanied by max pooling, denoted as:
 
 \begin{equation}
 \alpha(\cdot) = \textit{MAXPOOL}(\textit{Relu}(\textit{CONV}(\cdot)))
 \end{equation}
 
 Given the total time steps $T$ of the GGNN and the number of applications $l$ of $\alpha(\cdot)$, the $Conv$ module is represented as:
 
\begin{equation}
 Z_i^{1} = \alpha([H^{(T)}_{(v,g)}, x_i]), \ldots , Z_i^{(l)} = \alpha(Z_i^{(l-1)})
\end{equation}
\begin{equation}
Y_i^{(1)} = \alpha(H^{(T)}_{(v,g)}), \ldots, Y_i^{(l)} = \alpha(Y_i^{(l-1)})
\end{equation}
where we apply 1-D convolutional and dense layers to $[H^{(T)}_{(v,g)}, x_i]$ and $H^{(T)}_{(v,g)}$. Afterward, we make a pairwise multiplication on the two outputs and make a prediction. 

\begin{algorithm}[t]
\caption{collaborative distillation}
\KwIn{The parameters $\theta_1$ and $\theta_2$ of the student $S_1$ and $S_2$; The calculated local structure by LSP module; hyper-parameter $\alpha$ and label $y$}
\KwOut{$S_1$ and $S_2$ with excellent performance.}

Initialization parameters $\theta_1$ and $\theta_2$\;
\While{epochs $\leq$ max\_epoch}{
    \tcc{Training loss of $S_1$}
    Obtain the structure preserving loss $\mathcal{L}_{str}^{S_1}$ with Equ. (4) and Equ. (5)\;
    Obtain the cross-entropy loss $\mathcal{L}_{ce}^{S_1}(y, p_1)$\;
    Construct the total loss for $S_1$ as Equ. (7)\;
    \tcc{Training loss of $S_2$}
    Obtain the structure preserving loss $\mathcal{L}_{str}^{S_2}$ with Equ. (4) and Equ. (6)\;
    Obtain the cross-entropy loss: $\mathcal{L}_{ce}^{S_2}(y, p_2)$\;
    Construct the total loss for $S_2$ as Equ. (7)\;
    \tcc{Alternating training of $S_1$ and $S_2$}
    Update the parameter $\theta_1$ while keeping $\theta_2$ fixed\;
    Update the parameter $\theta_2$ while keeping $\theta_1$ fixed\;
}
\end{algorithm}

\subsection{Online Knowledge Distillation for GGNNs}
\blue{Knowledge Distillation (KD) is a technique in machine learning that transfers knowledge from a complex model (teacher) to a simpler model (student). Originally proposed for model compression, KD has evolved to become a powerful method for improving model performance across various domains. The methods of KD can be broadly categorized into three main approaches: 1) Soft Targets: This involves using the probability distributions over classes produced by the teacher model to train the student model to capture inter-class relationships . 2) Feature Embeddings: This approach focuses on transferring knowledge from the intermediate layers of the teacher model. It involves matching the feature representations or embeddings between the teacher and student models. This can be done through various methods such as minimizing the L2 distance between feature maps or using specially designed transformation layers. 3) Structural Knowledge: This method aims to transfer the structural information learned by the teacher model. In the context of neural networks, this could involve mimicking the attention patterns in transformer models or preserving the relational information in graph neural networks. Techniques like Local Structure Preservation (LSP) in graph neural networks fall under this category.}

Local Structure Preserving (LSP) is introduced as the first knowledge distillation module specifically designed for GNNs. Its purpose is to ensure that the local structure of the graph data learned by the student model closely resembles that of the teacher model, as referenced in \cite{yang2020distilling,guo2022alignahead}.
The method calculates the similarity \( L_{i,j} \) between each node \( i \) and its neighboring nodes \( j \), where \( (i,j) \in \epsilon \), in the feature space. This similarity is computed using one of several kernel functions and then normalized using a softmax function. The kernel functions for calculating the similarity $D(z_i, z_j)$ are given by the following equation:
\begin{equation}
D(z_i, z_j) = 
\begin{cases}
\|z_i - z_j\|_2^2 & \text{Euclidean} \\
z_i \cdot z_j & \text{Linear} \\
(z_i \cdot z_j + c)^d & \text{Poly} \\
e^{-\frac{1}{2\sigma}\|z_i-z_j\|^2} & \text{RBF}
\end{cases} 
\end{equation}

The similarity $L_{i,j}$ is further defined using a softmax-like normalization:

\begin{equation}
L_{i,j} = \frac{e^{D(z_i,z_j)}}{\sum_{j|(i,j)\in\epsilon} (e^{D(z_i,z_j)})},
\end{equation}

where $z_i$ and $z_j$ represent the node features. This equation normalizes the similarities, transforming them into a probability distribution.

The local structure $L_i$ of node $i $is described as the probability distribution of similarities between node $i$ and the nodes in its neighborhood.

To train the student model to emulate the local structure of the teacher model in the feature space, the Kullback-Leibler (KL) divergence is employed. The loss function for this LSP is given by:

\begin{equation}
\mathcal{L}_{lsp} = \frac{1}{N} \sum_{i=1}^N D_{KL} (L_i^T || L_i^S).
\end{equation}

LSP enables the transfer of local structural information from the teacher model to the student model. However, it requires a pre-trained teacher model. Inspired by this and \cite{guo2023online}, we use an idea of online knowledge distillation that does not require a pre-trained teacher model; instead, it relies on multiple student models learning from each other.

The core of this idea is a cross-layer knowledge distillation strategy, where one student layer is aligned with a layer at a different depth in another student model. During each round of alternate training, structure information from each student layer is transferred to the previous layer of another student model, eventually spreading the structure information across all layers.

In our work, we use two student models, $S_1$ and $S_2$, with identical GGNN architectures and $H$ intermediate layers. They use $l^{S_1}_{i,j}$ and $l^{S_2}_{i,j}$ to represent the local structure of node $j$ in the $i$-th layer of $S_1$ and $S_2$ respectively.

For $S_1$, the local structure of its $i$-th layer is required to match the $(i+1)$-th layer of $S_2$, with the final layer $H$ matching the first layer. The structure preserving loss is calculated as the sum of KL divergence losses over all layers:

\begin{equation}
\mathcal{L}_{str}^{S_1} = \sum_{i=1}^H L_{l_i}^{S_1}, \quad \mathcal{L}_{str}^{S_2} = \sum_{i=1}^H L_{l_i}^{S_2}
\end{equation}
where
\begin{equation}
\mathcal{L}_{l_i}^{S_1} = \frac{1}{N} \sum_{j=1}^N D_{KL} (l_{i+1,j}^{S_2} || l_{i,j}^{S_1})
\end{equation}

\begin{equation}
\mathcal{L}_{l_i}^{S_2} = \frac{1}{N} \sum_{j=1}^N D_{KL} (l_{i+1,j}^{S_1} || l_{i,j}^{S_2})
\end{equation}

The total loss of $S_1$ and $S_2$ is formulated as:

\begin{equation}
\mathcal{L}_1 = \mathcal{L}_{ce}^{S_1}(y, p_1) + \alpha L_{str}^{S_1}, \mathcal{L}_2 = \mathcal{L}_{ce}^{S_2}(y, p_2) + \alpha L_{str}^{S_2}
\end{equation}
where $\mathcal{L}_{ce}^{S_1/S_2}$ represents the cross entropy loss function, which is computed using the predictions ($p_1$ or $p_2$) from each student model and the true label $y$. The parameter $\alpha$ serves as a weighting factor, adjusting the relative importance of different loss components. As outlined in Algorithm 2, the training process alternates between the two student models, $S_1$ and $S_2$, allowing them to learn from each other iteratively. 

If the student number is greater than 2, each model will capture local structure information from all remaining models. The $i$-th layer’s structure preserving loss of the $k$-th model becomes
\begin{equation}
L^{S_k}_{l_i} = \frac{1}{M - 1} \frac{1}{N}\sum_{p=1 \vee p \neq k}^{M} \sum_{j=1}^{N} D_{KL}(l^{Sp}_{i+1,j} || l^{S_k}_{i,j})
\end{equation}
where $M$ is the number of student models. 

\blue{Distinct from prior endeavors, our online knowledge distillation procedure interlinks code language models with Graph Neural Networks (GNNs). Consequently, it achieves not only the transference of structural knowledge across GNN layers but also facilitates the amalgamation of semantic insights from language models within the GNN framework.}

\subsection{Extracting Code Sequence Semantics By CodeLMs}
\blue{CodeLMs are designed to understand, generate, and manipulate programming code. Similar to natural language models\cite{ni2021staresgru}, code language models are trained on vast corpora of source code from various programming languages and open-source repositories. They learn to recognize patterns, syntax, and semantics specific to programming languages, allowing them to perform a wide range of code-related tasks. Notable examples of code language models include: CodeBERT\cite{feng2020codebert}, GraphCodeBERT\cite{guo2020graphcodebert}, CodeT5\cite{wang2021codet5}, CuBERT\cite{kanade2020learning}, CodeGen\cite{nijkamp2022codegen} etc. These models typically use transformer-based architectures found in large language models. However, they are specifically fine-tuned on programming tasks and incorporate domain-specific knowledge about code structure and best practices. By learning semantic code representations from massive codebases, CodeLMs have great potential for identifying potential security vulnerabilities in source code.}

Now, we turn our attention to the handling of code sequences, which harbor a wealth of semantic information crucial for grasping the broader context and purpose of the code. Our approach to extracting this valuable semantic information begins with the transformation of function-level source code into code sequences. This conversion process is designed to preserve the logical flow that reflect the original order of declarations, operations, and control structures.

To process these code sequences, we fine-tune pre-trained codeLMs to capture the contextual information of various code elements, reveal implicit semantic features that may not be immediately apparent from the syntactic structure alone.  We select several representative code language models. CodeBERT, with its bimodal pre-training in programming and natural language, excels in understanding the relationship between code and its natural language descriptions. GraphCodeBERT extends CodeBERT by incorporating code structure information. CodeT5, as a unified pre-trained encoder-decoder model, offers versatility in both comprehension and generation tasks, which can be particularly beneficial for more complex semantic analyses. 

It's worth noting that our choice of code language model is deliberately flexible. We view these models as adaptable components in our larger framework, focusing on their utility in extracting semantic information rather than modifying the models themselves. This approach offers significant advantages: It allows easy updates as new, more advanced models become available.

\begin{figure*}
    \centering
    \includegraphics[width=\linewidth]{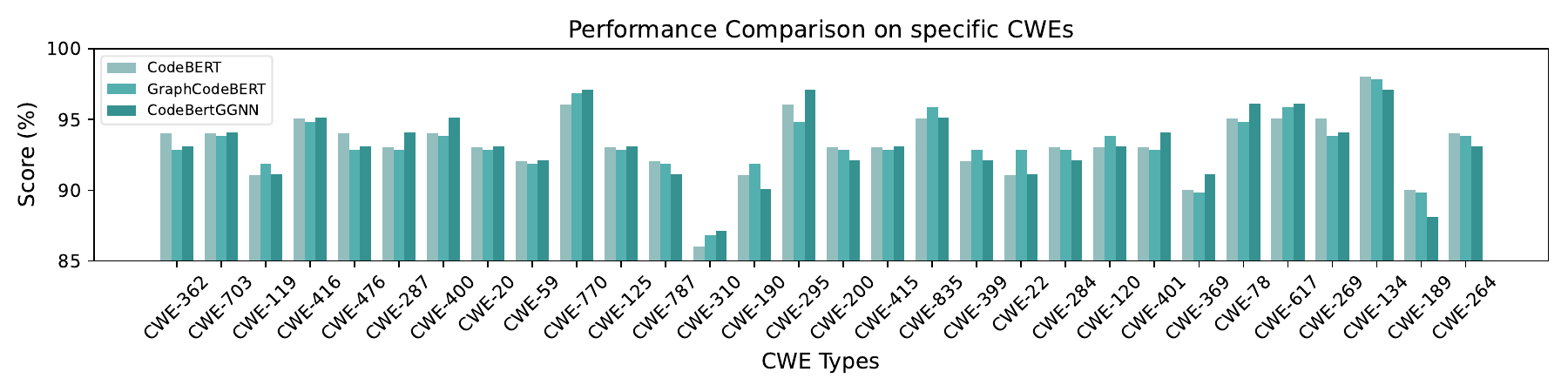}
    \caption{\textbf{Detection accuracy for the top-30 high-frequency CWE vulnerability types in DiverseVul. Note: CodeBertGGNN represents a basic variant of the proposed Vul-LMGNNs, specifically Vul-LMGNNs using CodeBERT as the Language Model (LM)}}
    \label{long}
\end{figure*}

\subsection{Joint Training of CodeLM and GNNs}
We propose an "implicit-explicit" joint training framework to synergistically optimize both CodeLM and GNNs, harnessing the strengths of both models. Our method unfolds in two distinct phases:

\begin{enumerate}
    \item Implicit Stage: We leverage CodeLM to initialize the node embeddings of the code GNN and simultaneously optimize both CodeLM and GNNs during the GNN's online knowledge distillation and label learning processes. This initialization process serves a dual purpose:
    
    \begin{enumerate}
        \item It facilitates the online knowledge distillation of the gated GNN, allowing for efficient transfer of knowledge from the language model to the graph structure.
        
        \item It enhances the label learning process, providing a rich semantic foundation for the GNN to build upon.
    \end{enumerate}
    
    By using CodeLM for initialization and continuous optimization, we ensure that the GNN starts with and maintains a strong semantic understanding of the code, which is then refined through its graph-based processing.

    \item Explicit Stage: This stage involves a more direct integration of the two models. We employ a linear interpolation technique to combine the outputs of CodeLM and the CodeLM-enhanced GNNs. This allows for a flexible, weighted combination of both models' predictions.
\end{enumerate}

The implicit phase ensures that the GNN benefits from CodeLM's deep language understanding from the outset, while the explicit phase allows for a nuanced integration of both models' strengths in the final output.

Next, we provide a detailed description of the two stages.

\subsubsection{Implicit Stage}
In our work, we utilize CodeLMs for initializing graph node embeddings. Specifically, we employ three different CodeLMs: CodeBERT\cite{feng2020codebert}, GraphCodeBERT, and CodeT5.For the sake of simplicity and clarity in our description, we will primarily focus on the process using CodeBERT as our exemplar CodeLM. The overall procedure, as depicted in Fig. \ref{feature}, involves the following steps.

We start by decomposing the function into a sequence of statement sets $C = {c_1, c_2, c_3, \ldots, c_n}$, where each $c_i$  is directly mapped to a node $v_i$  within the CPG. This mapping ensures that the complex structure of a function is represented as an interconnected graph of simpler, manageable elements. Each statement set is tokenized using CodeBERT's pretrained Byte Pair Encoding (BPE) tokenizer \cite{araabi2022effective}, converting the statement into a series of tokens.

Following this, we initialize the self-embedding layer weights using CodeBERT's trained word embeddings for each token and employ label encoding for node type embeddings. In parallel, efforts are made to fine-tune CodeBERT on our target dataset, intending to tailor the model’s understanding to our specific domain and thereby enhance the accuracy of the token vectorization process. Inspired by \cite{suneja2020learning}, we remove code properties from non-leaf nodes in the CPG, as these properties are often redundantly encoded in the leaf nodes. Finally,  the node content embeddings derived from CodeBERT and the node type embeddings obtained through label encoding are concatenated to form a comprehensive initial representation for each node.

In our joint training approach, we optimize the parameters of both CodeLM and GGNN, leveraging the complementary strengths of each model—CodeLM's contextual understanding and GGNN's relational insights—to improve the model's performance in detecting code vulnerabilities. This joint optimization strategy is implemented through the use of cross-entropy loss across code graph nodes, allowing for the simultaneous optimization of parameters for CodeLM and GGNN. The formulated loss function can be depicted as:

\begin{equation}
L = -\sum_{c=1}^{M} y_{ic} \log(\textit{Softmax}(\textit{MLP}(Z_i^{(l)}) \odot \textit{MLP}(Y_i^{(l)}))_{ic})
\end{equation}
$M$ represents the number of classes, $y_{ic}$ is a binary indicator (0 or 1) indicating whether class label $c$ is the correct classification for observation 
$i$.

In this training process, CodeLM updates the node embeddings with each iteration, thereby gradually improving the complementary advantages of both CodeLM and GGNN.
\subsubsection{Explicit Stage}
In the previous step, we implicitly combine CodeLM and GGNN by utilizing CodeLM to generate the node embeddings for GGNN. Here, we further explicitly combine the benefits of pre-training and graph-based approaches by leveraging interpolation predictions.
Specifically, we introduce an auxiliary classifier that operates directly on CodeLM embeddings by feeding code embeddings $E$ into a dense layer with softmax activation. Ultimately, we perform a linear interpolation \cite{lin2021bertgcn} of the predictions from Vul-LMGNNs and CodeLM, which is expressed as follows:
\begin{equation}
\textit{Pred} = \lambda \textit{Pred}_{\textit{GGCN}} + (1 - \lambda) \times \textit{Softmax(WE)}
\end{equation}

The parameter $\lambda$ controls the trade-off between the two objectives. A value of $\lambda = 1$ signifies the exclusive use of the full Vul-LMGNNs model, whereas $\lambda = 0$ indicates reliance solely on the CodeBERT module. When $\lambda$ is within the range $(0, 1)$, it allows for a balanced integration of the predictions from both models. The fine-tuned CodeLM model regulated and optimized the input graph for the GGNN. Subsequently, an interpolation prediction facilitated an appropriate trade-off between the graph model and sequence model detection results, yielding outstanding detection outcomes.

\begin{table}
  \centering
  \caption{\textbf{Summary of Datasets}}
  \label{tab:datasets}
  \small
  \resizebox{.475\textwidth}{!}{
  \begin{tabular}{@{}lllll@{}}
    \toprule
    \textbf{Dataset} & \textbf{\#Vulnerable} & \textbf{\#Non-Vul} &  \textbf{Source} & \textbf{CWEs}\\
    \midrule
    DiverseVul & 18,945 & 330,492 & Snyk, Bugzilla & 150 \\
    Devign & 11,888 & 14,149 & Github & N/A\\
    VDSIC & 82,411 & 119,1955 &GitHub, Debian & 4\\
    ReVeal & 1664 & 16,505 & Chrome, Debian & N/A\\
    \bottomrule
  \end{tabular}}
  \label{tab:dataset}
\end{table}

\section{Dataset Review}
\label{sec:dataset}
To evaluate our proposed code vulnerability detection method and other baseline methods, it is imperative to possess a substantial quantity of both vulnerable and non-vulnerable source code, spanning a diverse range of vulnerabilities.  In this paper, we have selected four public code vulnerability datasets, which include three widely-used popular datasets and one newly released comprehensive dataset. We have summarized the distribution of positive and negative samples and sources of the datasets, as well as whether they distinguish specific types of vulnerability, as shown in Table \ref{tab:dataset}.

The DiverseVul \cite{chen2023diversevul} dataset is a newly released dataset of vulnerable source code. It has been curated by crawling two security issue websites that feature the most commits in git systems, extracting commits that fix vulnerabilities and the corresponding source codes from the projects. The dataset also employs deduplication of functions based on their MD5 hashes. This dataset comprises 18,945 vulnerable functions spanning over 150 CWEs, and 330,492 non-vulnerable functions extracted from 7,514 commits. The range of projects covered by this dataset exceeds the total of all previous datasets by 295. This dataset’s substantial volume and diversity present a challenge for vulnerability detection methodologies.

The Devign dataset encompasses real-world function examples from GitHub, harvested from four renowned and diverse open-source libraries: Linux, FFmpeg, Qemu, and Wireshark. These examples are manually labeled based on commit messages and code differences. However, it does not provide information on the type of vulnerability or fine-grained labels. Additionally, this dataset is part of a programming language understanding evaluation benchmark known as CodeXGLUE \cite{lu2021codexglue}, and has been extensively used by various methods.

The Draper VDISC dataset \cite{russell2018automated} is an extensive collection of 1.27 million functions extracted from open-source software, annotated with insights from three distinct static analyzers to flag potential vulnerabilities. It encompasses the four most common CWEs: CWE-120, CWE-119, CWE-469, and CWE-476. Notably, the dataset exhibits a highly imbalanced distribution of positive and negative samples, with a ratio nearing 1:14.5. This imbalance could adversely affect the real performance of our testing models. Therefore, in this paper, we have utilized a pre-processed version of the dataset with a more balanced distribution.

REVEAL\cite{chakraborty2021deep} is a comprehensive real-world dataset, amassed by monitoring historical vulnerabilities from two prominent open-source projects: the Linux Debian Kernel and Chromium.  It involves the extraction of the respective vulnerable and fixed versions of C/C++ source and header files that have been modified in patches, serving as positive and negative samples for research.

\section{Experiments results}
\label{sec:results}
In this section, we present the experimental setup and the outcomes of our evaluations conducted on our proposed model, alongside six state-of-the-art baselines across four datasets. We have formulated the following four Research Questions (RQs) and have addressed them through our experimental investigations:

\begin{itemize}[left=2pt,label=$\bullet$,itemsep=2pt]
\item \textbf{RQ1:} How does our Vul-LMGNNs performance compare with other learning-based methods for vulnerability identification?

\item \textbf{RQ2:} With the variation of the trade-off parameter, what changes can be observed in the model’s performance?

\item \textbf{RQ3:} Is the fine-tuning process of pre-trained models a more efficient method for token vectorization in node embedding compared to initial word embedding weights?

\item \textbf{RQ4:} How do different GNN architectures and pre-trained models influence the overall performance of the model?

\item \textbf{RQ5:} Does online knowledge distillation enhance model performance? How does the number of student models alter performance?
\end{itemize}

The experiments were executed on a single NVIDIA A100 80GB GPU. The system specifications comprised NVIDIA driver version 525.85.12 and CUDA version 11.8. The software environment was configured with Python 3.10.13 and torch 2.2.0.

\begin{table*}
\centering
\caption{\textbf{Performance metrics of various models on datasets with specific CWEs.Note: C-B: CodeBERT, GC-B: GraphCodeBERT, C-T5s: CodeT5-Small, C-T5b: CodeT5-Base.}}
\label{tab:baseline_2}
\renewcommand{\arraystretch}{1.2}
\footnotesize 
\resizebox{1.02\textwidth}{!}{
\begin{tabular}{p{0.5\textwidth}p{0.5\textwidth}}
\begin{tabular}{l|cccc}
\hline
\multicolumn{5}{c}{\textbf{DiverseVul Dataset}} \\
\hline
\textbf{Model} & \textbf{ACC (\%)}  & \textbf{P (\%)} & \textbf{R (\%)} & \textbf{F1 (\%)} \\ \hline
Bert      & 91.99  & 27.95 & 13.09 & 17.83\\ 
CodeBert          & 92.40 & 28.26 & 20.02 & 23.44\\ 
GraphCodeBert     & 92.96& 31.14& 16.30& 21.40\\
CodeT5-Small     & 94.27& 43.00& 17.46& 24.83\\
CodeT5-Base     & 94.24& 42.96& 19.05& 26.40\\
TextCNN   & 92.16& 10.25& 9.82& 10.03\\ 
TextGCN      & 91.50& 15.66& 11.50& 13.27\\
\hline
Devign(AST)           & 70.21& 9.35& 9.22& 9.28\\
VulBERTa-MLP          & 92.14& 21.47& 16.92& 18.92\\
VulBERTa-CNN          & 91.33& 17.76& 16.51& 17.11\\
TokenCNN          & 90.54& 11.21& 10.76& 10.98\\
VulDeePecker          & 91.76& 17.31& 13.76& 15.33\\
SySeVR          & 92.86& 27.10& 18.76& 22.17\\
CFExplainer          & 88.86& 13.63& 19.76& 16.13\\
ReGVD          & 85.86& 10.26& 20.16& 13.73\\
VulMAE          & 88.78& 11.21& 15.46& 13.00\\
ReVeal          & 85.88& 9.35& 18.46& 12.42\\ 
\hline
Vul-LMGNNs(C-B)         & \textbf{93.06}& \textbf{32.21} & \textbf{18.54}& \textbf{23.54}\\
Vul-LMGNNs(GC-B)        & \textbf{93.61}& \textbf{33.73} & \textbf{18.51}& \textbf{23.90}\\
Vul-LMGNNs(C-T5s)        & \textbf{95.04}& \textbf{60.51} & \textbf{20.01}& \textbf{30.43}\\ 
Vul-LMGNNs(C-T5b)        & \textbf{94.84}& \textbf{55.86} & \textbf{23.01}& \textbf{32.59}\\ 
\hline
\end{tabular}
&
\begin{tabular}{l|cccc}
\hline
\multicolumn{5}{c}{\textbf{Draper VDSIC Dataset}} \\
\hline
\textbf{Model} & \textbf{ACC (\%)}  & \textbf{P (\%)} & \textbf{R (\%)} & \textbf{F1 (\%)} \\ \hline
Bert      &79.41  &81.86 &75.97 &78.80\\ 
CodeBert          &83.13 &86.13 &78.97 & 82.39\\ 
GraphCodeBert     &83.98& 84.74& 83.17& 83.95\\
CodeT5-Small     & 84.47& 84.21& 84.70& 84.51\\
CodeT5-Base     & 85.01& 85.90& 84.64& 85.12\\
TextCNN   & 66.54& 65.36& 70.55& 67.86\\ 
TextGCN      & 67.55& 67.66& 67.63& 67.64\\
\hline
Devign(AST)           &59.30 & 58.84& 68.93 & 63.49\\ 
VulBERTa-MLP          & 80.41& 83.37& 75.97& 79.50\\
VulBERTa-CNN          & 78.98& 79.46& 78.17& 78.81\\
TokenCNN          & 65.05& 63.99& 68.84& 66.33\\
VulDeePecker          & 78.15& 78.50& 77.53& 78.01\\
SySeVR          & 82.01& 83.25& 80.15& 81.67\\
CFExplainer          & 80.01& 83.37& 74.97& 78.95\\
ReGVD          & 79.78& 80.15& 79.17& 79.66\\
VulMAE          & 77.05& 76.12& 78.84& 77.45\\
ReVeal          & 78.15& 78.50& 77.53& 78.01\\ 
\hline
Vul-LMGNNs(C-B)       & \textbf{84.38}& \textbf{87.37} & \textbf{80.64}& \textbf{83.87} \\
Vul-LMGNNs(GC-B)        & \textbf{84.51}& \textbf{86.36} & \textbf{81.97}& \textbf{84.11}\\
Vul-LMGNNs(C-T5s)        & \textbf{85.68}& \textbf{86.69} & \textbf{84.31}& \textbf{85.48}\\ 
Vul-LMGNNs(C-T5b)        & \textbf{85.91}& \textbf{86.33} & \textbf{85.33}& \textbf{85.83}\\ 
\hline
\end{tabular}
\end{tabular}}
\end{table*}

In our comparative analysis, we evaluated the performance of Vul-LMGNN against 17 state-of-the-art detection models. This includes 5 Transformer-based models: BERT \cite{devlin2018bert}, CodeBERT \cite{feng2020codebert}, GraphCodeBERT \cite{guo2020graphcodebert}, CodeT5-Small \cite{wang2021codet5}, and CodeT5-Base \cite{wang2021codet5}. We also included 2 classical text-based techniques commonly used for vulnerability detection: TextGCN \cite{yao2019graph} and TextCNN \cite{guo2019improving}. In addition, we evaluated 10 specialized vulnerability detection algorithms: Devign \cite{zhou2019devign}, VulBERTa-MLP \cite{hanif2022vulberta}, VulBERTa-CNN \cite{hanif2022vulberta}, TokenCNN \cite{russell2018automated}, VulDeePecker \cite{li2018vuldeepecker}, SySeVR \cite{li2021sysevr}, CFExplainer \cite{yang2024cfexplainer}, ReGVD \cite{nguyen2022regvd}, VulMAE \cite{zamani2023vulmae}, and ReVEAL \cite{chakraborty2021deep}. For computational efficiency, functions with a node size exceeding 500 in the CPG were excluded from our analysis. In terms of our model’s configuration, the learning rate and batch size were set to $1e-4$ and 64, respectively. The training was conducted over 20 epochs with an early stopping criterion triggered if no further optimization in performance. Specifically for the Devign model, AST was employed for the code graph representation. 

\blue{To ensure fairness, we used official code repositories for baseline models when available and followed original papers closely where not. All models were trained under identical conditions, including hardware, dataset splits, batch sizes, and optimizers. Key parameters were set according to original recommendations or through grid search.}

\begin{table*}
\centering
\caption{\textbf{Performance metrics of various models on datasets with no specific CWEs.Note: C-B: CodeBERT, GC-B: GraphCodeBERT, C-T5s: CodeT5-Small, C-T5b: CodeT5-Base.}}
\label{tab:baseline_1}
\footnotesize 
\renewcommand{\arraystretch}{1.2}
\resizebox{1.02\textwidth}{!}{
\begin{tabular}{p{0.5\textwidth}p{0.5\textwidth}}
\begin{tabular}{l|cccc}
\hline
\multicolumn{5}{c}{\textbf{Devign Dataset}} \\
\hline
\textbf{Model} & \textbf{ACC (\%)}  & \textbf{P (\%)} & \textbf{R (\%)} & \textbf{F1 (\%)} \\ \hline
Bert      & 60.58  & 57.67 & 54.64 & 56.11\\ 
CodeBert          & 63.93 & 61.03 & 58.10 & 59.53\\ 
GraphCodeBert     & 64.80& 64.37& 54.38& 58.96\\
CodeT5-Small     & 65.62& 64.35& 55.39& 59.53\\
CodeT5-Base     & 65.92& 64.27& 57.12& 60.48\\
TextCNN   & 60.38& 59.03& 57.72& 58.37\\ 
TextGCN      & 60.47& 60.87& 58.58& 59.70\\ 
\hline
Devign(AST)           & 57.66& 56.96& 56.25& 56.60\\
VulBERTa-MLP          & 64.75& 62.71& 56.22& 59.29\\
VulBERTa-CNN          & 64.42& 63.11& 53.12& 57.69\\
TokenCNN          & 47.63& 43.98& 45.01& 43.97\\
VulDeePecker          & 60.36& 56.94& 55.41& 56.39\\
SySeVR          & 62.18& 60.45& 54.05& 57.30\\
CFExplainer          & 57.66& 56.96& 56.25& 56.60\\
ReGVD          & 61.89& 60.74& 48.20& 53.75\\
VulMAE          & 62.36& 60.47& 50.71& 55.16\\
ReVeal          & 62.38& 59.80& 53.71& 56.59\\
\hline
Vul-LMGNNs(C-B)        & \textbf{65.70}& \textbf{64.53} & \textbf{56.34}& \textbf{60.16}\\ 
Vul-LMGNNs(GC-B)        & \textbf{66.33}& \textbf{64.73} & \textbf{57.77}& \textbf{61.01}\\
Vul-LMGNNs(C-T5s)        & \textbf{66.77}& \textbf{63.45} & \textbf{64.20}& \textbf{63.82}\\ 
Vul-LMGNNs(C-T5b)        & \textbf{67.27}& \textbf{64.73} & \textbf{62.20}& \textbf{63.44}\\ 
\hline
\end{tabular}
&
\begin{tabular}{l|cccc}
\hline
\multicolumn{5}{c}{\textbf{ReVeal Dataset}} \\
\hline
\textbf{Model} & \textbf{ACC (\%)}  & \textbf{P (\%)} & \textbf{R (\%)} & \textbf{F1 (\%)} \\ \hline
Bert      & 86.88  & 32.70 & 40.13 & 36.04\\ 
CodeBert          & 88.64 & 38.26 & 38.13 & 38.19\\ 
GraphCodeBert     & 89.25& 41.67& 41.81& 41.74\\
CodeT5-Small     & 90.50& 47.81& 40.73& 43.99\\
CodeT5-Base     & 90.94& 50.69& 39.76& 44.56\\
TextCNN   & 85.43& 26.32& 20.33& 22.94\\ 
TextGCN      & 87.25& 24.61& 17.85& 20.69\\
\hline
Devign(AST)           & 87.49 & 36.65& 31.55 & 33.91\\ 
VulBERTa-MLP          & 88.48& 36.79& 35.90& 36.34\\
VulBERTa-CNN          & 87.64& 34.46& 38.76& 36.48\\
TokenCNN          & 73.48& 17.47& 50.90& 26.01\\
VulDeePecker          & 89.05& 17.68& 13.87& 15.70\\
SySeVR          & 84.22& 24.36& 40.11& 30.25\\
CFExplainer          & 85.18& 24.21& 29.01& 26.39\\
ReGVD          & 90.63& 64.70& 14.47& 23.65\\
VulMAE          & 88.24& 30.25& 21.76& 25.31\\
ReVeal          & 85.37& 29.90& 40.91& 33.87\\
\hline
Vul-LMGNNs(C-B)        & \textbf{90.80}& \textbf{57.09} & \textbf{46.45}& \textbf{51.22}\\ 
Vul-LMGNNs(GC-B)        & \textbf{91.58}& \textbf{55.12} & \textbf{43.41}& \textbf{48.57}\\
Vul-LMGNNs(C-T5s)        & \textbf{90.98}& \textbf{50.76} & \textbf{50.41}& \textbf{50.58}\\ 
Vul-LMGNNs(C-T5b)        & \textbf{91.68}& \textbf{54.89} & \textbf{51.41}& \textbf{53.09}\\ 
\hline
\end{tabular}
\end{tabular}}
\end{table*}

\subsection{Comparison with Baselines \textbf{(RQ1)}}
To evaluate the performance of Vul-LMGNNs on code vulnerability detection,  we executed an extensive comparative analysis against 17 baseline models utilizing the four datasets delineated in Table \ref{tab:dataset}. The experimental results are systematically presented in Table \ref{tab:baseline_2} and \ref{tab:baseline_1}.

We initially tested the Vul-LMGNNs on datasets categorized by specific CWEs and analyzed \blue{their} capability to recognize these CWEs within the test set. In terms of accuracy, precision, and F1 score, Vul-LMGNNs outperformed all baseline models. Specifically, any variant of Vul-LMGNNs outperformed all baseline models. Specifically, on the DiverseVul dataset, Vul-LMGNNs with CodeBERT as a component achieved an accuracy of 93.06\% and an F1 score of 23.54\%; on the balanced version of the VDSIC dataset, \blue{they} reached an accuracy of 84.38\%. Vul-LMGNNs with CodeT5 as a component achieved an accuracy of 95.04\% on the DiverseVul dataset, with an F1 score of 32.59\%, surpassing the previous best method by 10\%.

\blue{Vul-LMGNNs outperform other models primarily due to their effective integration of structural information from GNNs and semantic understanding from pre-trained language models. While models like BERT and CodeBERT capture sequential semantics well, they lack the structural insights needed for vulnerability detection in complex codebases. In contrast, graph-based models like TextGCN struggle in our experiments, as they focus on word co-occurrence and miss critical code structure details. By leveraging both semantic and structural features, Vul-LMGNNs achieve superior performance across various metrics, excelling in both precision and recall when detecting vulnerabilities embedded in the code.}

Furthermore, extensive experiments demonstrated that code language models trained on massive datasets generally outperform specialized vulnerability detection methods, despite not containing C/C++ programs in their pre-training datasets. Pre-trained LM like BERT, CodeBERT, and CodeT5 variants consistently achieved higher accuracy (91.99\% - 94.27\%) compared to specialized methods such as VulDeePecker (91.76\%), SySeVR (92.86\%), and ReVeal (85.88\%). CodeT5-Small and CodeT5-Base demonstrated particularly strong performance, benefiting from their larger architectural scale and pre-training, with accuracies of 94.27\% and 94.24\% respectively, and F1 scores of 24.83\% and 26.40\%. The other three datasets exhibitd similar trends. These results strongly suggest that the transfer learning capabilities and rich semantic understanding of large-scale pre-trained code language models provide a significant advantage in vulnerability detection tasks. Their ability to capture complex code patterns and contextualized representations appears to be more effective than the feature engineering approaches typically used in specialized vulnerability detection methods.

As shown in Figure \ref{long}, we demonstrated the performance of our method on specific vulnerabilities. We compared the competitiveness of our approach with CodeBert and GraphCodeBert. In this comparison, our method is the Vul-LMGNNs with CodeBert as a component, which we refer to as CodeBertGGNN. Among the top 30 most frequently occurring CWEs in the test set, our model achieved \blue{the highest accuracy rate on 50\% of CWEs}. It can be observed that for some CWEs, the recognition accuracy of the model is generally low, such as CWE-310 (Cryptographic Issues) and CWE-189 (Numeric Errors), while for another subset of CWEs, there are high recognition accuracy rates, such as CWE-134 (Controlled Format String) and CWE-770 (Allocation of Resources Without Limits or Throttling). 

\blue{We analyzed the lower identification rates for CWE-310 and CWE-189 and identified several potential reasons for these challenges:

Regarding CWE-310 (Cryptographic Issues):
These vulnerabilities primarily involve encryption problems, which can manifest in cryptographic algorithms, key management, or protocol implementations within the code. Vulnerability detection tools face difficulties in identifying CWE-310 class vulnerabilities for the following reasons: 1). While the model can learn some indicators of cryptographic vulnerabilities from contextual dependencies, certain encryption issues only become apparent under specific runtime conditions. 2). The complexity and subtlety of cryptographic implementations often require domain-specific knowledge that may be beyond the scope of general-purpose vulnerability detection tools.

Concerning CWE-189 (Numeric Errors):
The challenges in detecting these vulnerabilities are similar to those of CWE-310: 1). Some numeric errors only manifest under specific input conditions or runtime environments, making it difficult for automated tools to capture these dynamic behaviors. 2). There is a significant human factor involved in CWE-189 vulnerabilities. Developers may introduce counting errors due to mistakes or oversights during coding, which are often subtle and not easily detectable by automated tools.

In both cases, while our model can infer some vulnerability indicators from code context and structure, it may struggle with issues that only become apparent during runtime or require deep domain-specific understanding. }

\begin{figure}[t]
    \centering
    \includegraphics[width=0.75\linewidth]{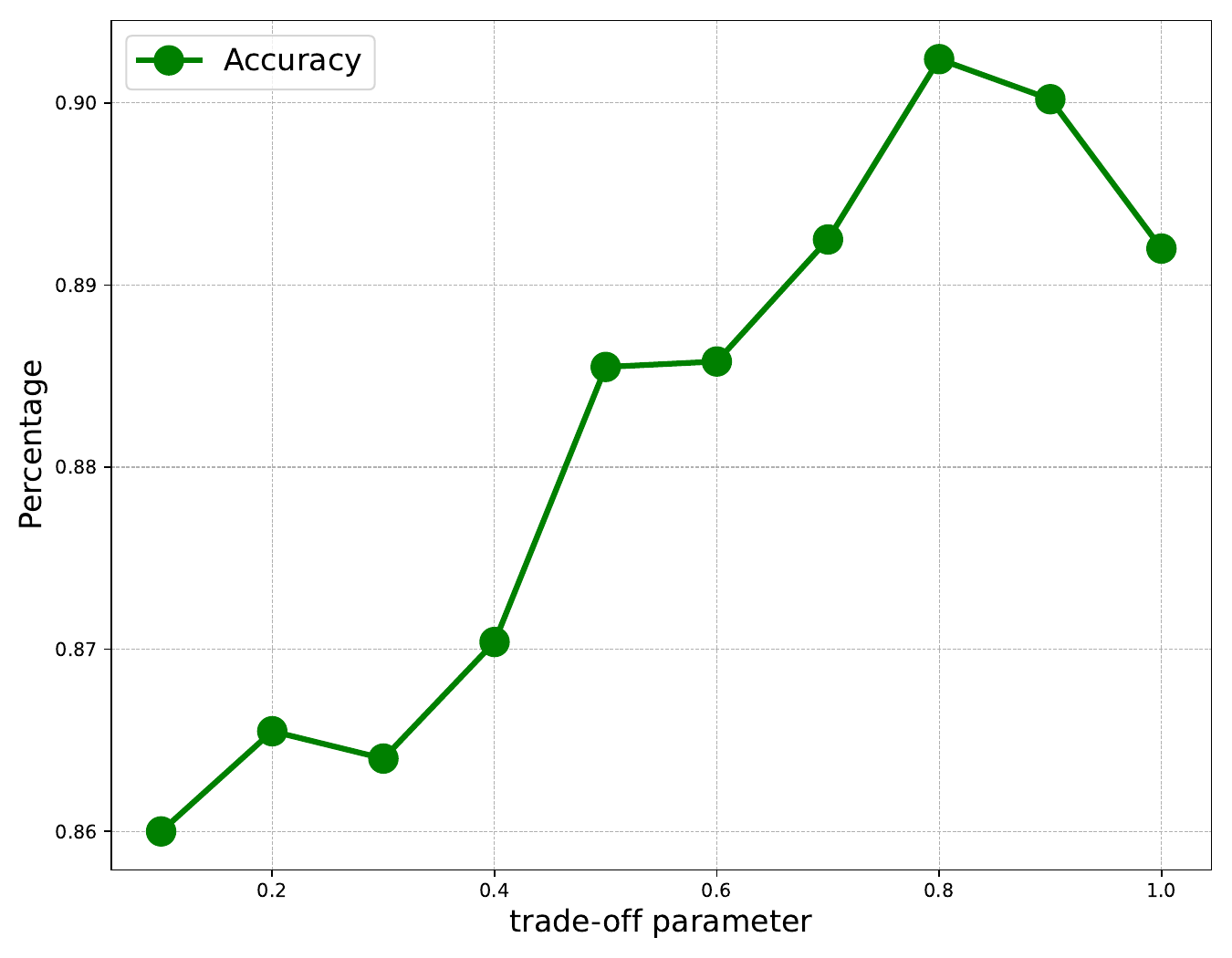} 
    \caption{\textbf{Accuracy of Vul-LMGNNs when varying trade-off parameter on partial DiverseVul dataset.}}
    \label{trade-off_1}
\end{figure}

\begin{figure}[t]
    \centering
    \includegraphics[width=0.78\linewidth]{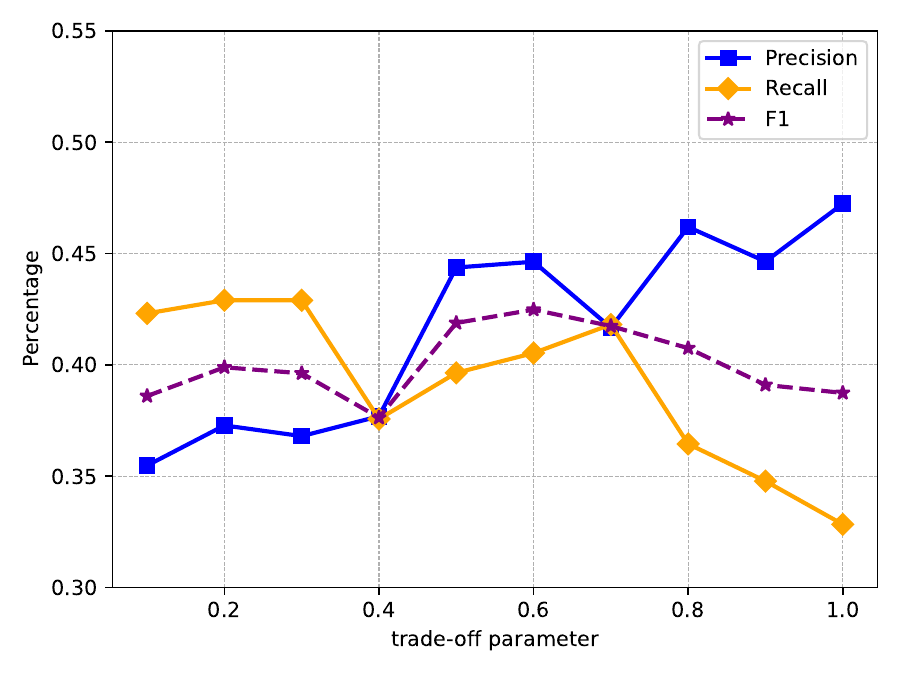}
    \caption{\textbf{Precision, recall and f1 of Vul-LMGNNs when varying trade-off parameter on partial DiverseVul dataset.}}
    \label{trade-off_2}
\end{figure}

In the realm of graph-based detection models, TextGCN, while performing well in text classification, showed mediocre results in code vulnerability detection experiments, achieving only a 91.50\% accuracy rate on DiverseVul, with precision and recall at 15.66\% and 11.50\%, respectively. This may be due to TextGCN’s focus on word co-occurrence, lacking structural code information. The AST version of Devign, which incorporates control flow and data flow information and uses Word2Vec along with the average of tokens for node vector representation, performed poorly with an accuracy of only 70.21\%, a gap of 14.26\% - 23.31\% from our F1 score. This could be attributed to the neglect of local semantic information of code within the node. 

\blue{In vulnerability detection, false negatives (missed vulnerabilities) pose a significant risk. Vul-LMGNNs demonstrate strong recall across datasets, reducing false negatives compared to baseline models like Devign and ReVeal. On the ReVeal dataset, our Vul-LMGNN model achieved a recall of 51.41\%, significantly higher than ReVeal's 40.91\%. In security applications, lowering false negatives is critical, and our model's enhanced recall showcases its robustness in identifying vulnerabilities that other methods might miss. Additionally, we observed that false positives, although slightly higher in our model, are a reasonable trade-off for the improved vulnerability detection coverage.}

The performance disparity on the other two datasets without specific CWEs is similar to those observed in the previous CWE-specific evaluations, as illustrated in Table \ref{tab:baseline_1}. our proposed Vul-LMGNNs models across both the Devign and ReVeal datasets, showcasing significant improvements over existing approaches in vulnerability detection tasks. On the Devign dataset, Vul-LMGNNs(C-T5b) variant achieved the highest scores, with an accuracy of 67.27\%, precision of 64.73\%, recall of 62.20\%, and F1 score of 63.44\%.  The performance gap is even more pronounced on the ReVeal dataset. Here, Vul-LMGNNs(C-T5b) attained remarkable results with an accuracy of 91.68\%, precision of 54.89\%, recall of 51.41\%, and F1 score of 53.09\%.  The performance disparity between Vul-LMGNNs and other models is particularly evident when comparing to traditional approaches. For instance, on the ReVeal dataset, the best performing Vul-LMGNNs model achieved an F1 score that is nearly 17 percentage points higher than best baseline  (53.09\% vs. 36.48\%).

In summary, the experimental results provide strong evidence for the effectiveness of Vul-LMGNNs in vulnerability detection tasks. \blue{The models' ability to capture both structural and semantic information in code results in improved recall, lower false negatives, and consistent outperformance across datasets.} The models' ability to outperform existing state-of-the-art approaches across multiple metrics and datasets underscores their potential to significantly advance the field of automated code vulnerability detection.

\subsection{Impact of the Tradeoff Parameter \textbf{(RQ2)}}
The parameter $\lambda$ controls the trade-off between the training CodeLM and GNN. As $\lambda$ approaches 1, the model’s decisions rely more on the graph structure with the PL model embedding layer. Conversely, when $\lambda$ approaches 0, the model leans toward sequence-based decisions. Our experiments using Vul-LMGNNs (C-B) across various datasets indicated that the optimal value of $\lambda$ varies for different tasks, likely due to variations in vulnerability types and data distributions. For instance, on the partial Draper VDSIC dataset, increasing $\lambda$ does not significantly improve model performance. This phenomenon can be attributed to the strong performance of sequence-based methods on the former VDSIC dataset. 

Fig. \ref{trade-off_1} and \ref{trade-off_2} illustrated the evaluation matrix of Vul-LMGNNs with varying $\lambda$ on the partial DiverseVul dataset. Accuracy improves consistently as $\lambda$ increases, reaching its peak at $\lambda$=0.8, slightly outperforming the GGNN or CodeBert alone (at $\lambda$ = 0 or 1). The achieved accuracy is 90.24\%. During this process, precision exhibits fluctuations but overall shows an upward trend, reaching 46. 19\% at $\lambda$ = 0.8, an improvement of 10. 71\% over using the sequence model alone. However, recall follows a declining trend, reaching 36.45\% at $\lambda$=0.8, indicating that transformer-based PL models exhibit higher recall in certain vulnerability detection scenarios.

\begin{table}[t]
\centering
\caption{\textbf{Performance across various node embedding and initialization methods.}}
\setlength{\tabcolsep}{9pt} 
\renewcommand{\arraystretch}{1.2}
\resizebox{0.5\textwidth}{!}{
\begin{tabular}{lcccc}
\toprule
 \textit{\textbf{Base}} & \textbf{ACC(\%)} & \textbf{P(\%)} & \textbf{R(\%)} & \textbf{F1(\%)} \\
\midrule
CodeBERT & 84.35 & \textbf{87.75} & 79.85  & 83.61 \\
GraphCodeBert & 84.05 & 86.46 & 80.74& 83.50 \\ 
Fine-tuned(C-B) & 84.38 & 87.37 & 80.64 & 83.87 \\
Fine-tuned(GC-B) & \textbf{84.51} & 86.36 & \textbf{81.97} & \textbf{84.11} \\
\bottomrule
\end{tabular}}
\label{tab:fine-tune}
\end{table}

\subsection{Evaluation of Fine-tuning Process  \textbf{(RQ3)}}
Pre-trained LMs demonstrate outstanding performance in various natural language processing tasks. Currently, there is a growing focus among researchers on employing pre-trained LMs for code-related tasks, including code search, code completion, code summarization and so on \cite{tang2023csgvd,wang2021codet5,chakraborty2022natgen,xu2022systematic}. This has led to promising results in applications. This has prompted us to incorporate pre-trained models for programming languages in order to construct a novel vulnerability detection model.

We utilized the word embedding layer of pre-trained models as tokenization tools to generate node embeddings for graphs. These embeddings' weights are further fine-tuned during training. In our experiments, we explored three settings. First, we initialized our embedding layer weights using a fine-tuned CodeBERT which perform fine-tuning on the target dataset \cite{10.1145/3597926.3598036}. Additionally, we compared this approach with two others: initializing embedding layer weights directly using pre-trained CodeBERT and GraphCodeBERT, respectively. The results are summarized in Table \ref{tab:fine-tune}.

The experimental results revealed that fine-tuning Code LMs models improves performance over using the base models for node embedding initialization. the base model CodeBERT achieved an accuracy of 84.35\%, with an F1 score of 83.61\%. When employing GraphCodeBERT, the accuracy slightly decreased to 84.05\%, while F1 score improved to 83.50\%. In contrast, fine-tuned CodeBERT (C-B) model achieved an accuracy of 84.38\%, an F1 score of 83.87\%. Notably, the fine-tuned GraphCodeBERT (GC-B) model outperformed all others with an accuracy of 84.51\%, and an F1 score of 84.11\%. These results suggested that fine-tuning the models leads to a better overall balance in performance metrics.

\subsection{Different GNN Model Combination \textbf{(RQ4)}}
To investigate the impact of combining different GNNs with pre-trained LMs on vulnerability detection tasks, we compared distinct GNNs: Graph Gated Neural Network (GGNN), Graph Convolutional Network (GCN) \cite{kipf2016semi}, and Graph Attention Network (GAT) \cite{velivckovic2017graph}, and three variants of GraphSAGE (SAGE-mean, SAGE-GCN and SAGE-pool), all integrated with CodeBERT. For a fair comparison, we followed the configuration from \cite{tang2023csgvd}, maintaining a consistent three-layer GNN architecture and setting GAT’s number of heads to 8. Additionally, we employed fine-tuned CodeBERT with consistent model parameters. The specific experimental results are shown in Table \ref{tab:gnn}.

\begin{table}[t]
\centering
\caption{\textbf{Performance across various GNN architectures.}}
\setlength{\tabcolsep}{8.5pt} 
\renewcommand{\arraystretch}{1.2}
\begin{tabular}{lcccc}
\toprule
 \textit{\textbf{Combination}} & \textbf{ACC(\%)} & \textbf{P(\%)} & \textbf{R(\%)} & \textbf{F1(\%)} \\
\midrule
GGNN+\\CodeBERT & \textbf{84.38} & \textbf{87.37}& \textbf{80.64}  & \textbf{83.87} \\
GCN+\\CodeBERT & 83.08 & 86.90 & 77.90& 82.15 \\
GAT+\\CodeBERT & 79.29 & 81.92 & 75.15 & 78.39 \\
SAGE-mean+\\CodeBERT & 83.55 & 86.93 & 78.97 & 82.76 \\
SAGE-GCN+\\CodeBERT & 84.18 & 87.05 & 80.31 & 83.54 \\
SAGE-pool+\\CodeBERT & 81.90 & 85.82 & 76.43 & 80.85 \\
\bottomrule
\end{tabular}
\label{tab:gnn}
\end{table}

As shown in Table 5, our experiments were conducted on the partial Draper VDSIC dataset. The results indicate that GGNN exhibit the best overall performance, with an accuracy of 84.38\% and an F1 score of 83.87\%. Compared to GGNN, GCN experiences decreases in accuracy and precision by 1.3\% and 0.47\%, respectively, with the most significant decrease observed in recall at 2.74\%. This may be attributed to GCN treating all neighboring nodes equally during convolution, thus failing to assign different weights based on node importance, leading to inaccurate identification of nodes related to code vulnerabilities. Additionally, GCN updates node features for the entire graph in a single computation, which poses challenges when dealing with complex code graph structures in inductive learning tasks related to code vulnerabilities.

The performance of the GAT model exhibited a considerable gap compared to the previous two, with an accuracy of only 79.29\%. Although GAT utilizes self-attention mechanisms to represent each node as a weighted sum of its neighbors, it does not fully leverage edge information, only utilizing connectivity, whereas edge information encompasses the control and data flow information of the code. In contrast, GGNN employs GRU units, allowing each node to receive messages from neighboring nodes at each iteration. This approach effectively captures both code data flow features and long sequence dependencies, resulting in outstanding performance.

\begin{table}[t]
\centering
\caption{\textbf{Ablation study for online knowledge distillation; "self" denotes the absence of online knowledge distillation.}}
\setlength{\tabcolsep}{8.5pt} 
\renewcommand{\arraystretch}{1.2}
\begin{tabular}{lcccc}
\toprule
 \textit{\textbf{KD Ablation}} & \textbf{ACC(\%)} & \textbf{P(\%)} & \textbf{R(\%)} & \textbf{F1(\%)} \\
\midrule
Self & 83.50 & 86.68& 79.17  & 82.75 \\
2 Student & \textbf{84.38} & 87.37& 80.64  & 83.87 \\
3 Student & 84.18 & 84.77 & \textbf{83.33} & \textbf{84.04} \\
4 Student & 83.90 & 87.93 & 78.59 & 83.00 \\
5 Student & 84.15 & \textbf{88.93} & 78.01 & 83.11 \\
6 Student & 83.61 & 86.22 & 80.01 & 83.00 \\
\bottomrule
\end{tabular}
\label{tab:KG}
\end{table}

\subsection{Evaluation of Online KD for GGNNs \textbf{(RQ5)}}
Figure \ref{KD} reveals the flow of structural information during the iterative training process of two GNNs in the online knowledge distillation approach. It can be observed that each layer of one student model aligns with the next layer of the other student model, with the final layer particularly matching the first layer. In this way, the two student models exchange structural information with each other, and propagate it across different layer depths. In this work, we evaluated two aspects of this approach: 1) the performance gain achieved and 2) how the performance changes as the number of student models is varied. 

When transitioning from the "Self" baseline, which denotes the absence of online knowledge distillation, to using two student models, there is a significant enhancement across all measured metrics. This initial improvement underscores the benefit of the distillation process. Further analysis shows that the performance continues to improve with the addition of a third student model. However, this trend does not persist uniformly as the number of student models increases beyond three. For instance, while the metrics improve when moving from two to three student models, the subsequent addition of a fourth student model does not yield substantial gains and is followed by a slight decline in Accuracy and Precision. This indicates that, after reaching a certain threshold, the benefits of adding more student models begin to diminish.

\subsection{Compared Online KD with MixHop}
\blue{In this paper, we leverage Online KD to address the limitations of single-layer information propagation in GNNs. This approach enables our method to effectively capture the intricate logic within abstract code graphs. While acknowledging the existence of algorithms like MixHop\cite{abu2019mixhop} in the GNN literature, which aim to capture long-range dependencies in graph neural networks, our experiment seeks to analyze and compare our KD-based method with MixHop.

We begin by examining the fundamental differences between these two approaches. MixHop is designed to overcome GNNs' limitations in handling long-distance relationships between nodes by mixing node features across multiple hop counts (i.e., neighborhood layers). In contrast, our Online Knowledge Distillation method focuses on transferring knowledge between teacher and student models. The key distinction is that online KD facilitates knowledge transfer between teacher and student models, whereas MixHop integrates node features across multiple hop counts. The implementation results are presented in the Table\ref{tab:comparison}, providing a comparative analysis of these two approaches in addressing the challenges of information propagation and logic capture in GNNs for abstract code graphs.

Based on the experimental results, the Knowledge Distillation (KD) based method demonstrates superior performance compared to the MixHop-based approach on the Partial Draper VDSIC Dataset. The KD-based method achieves higher accuracy (84.38\% vs 83.68\%) and significantly better recall (80.64\% vs 78.47\%), indicating its enhanced ability to identify relevant instances. The KD-based approach's higher F1 score (83.87\% vs 82.78\%) underscores its better overall balance between precision and recall. These results suggest that the KD-based method is more effective at capturing and utilizing relevant information from the dataset, making it the preferable choice.}

\begin{table}[t]
\centering
\caption{Comparison of Experimental Results for MixHop-based Methods and KD-based Methods on the Partial Draper VDSIC Dataset}
\begin{tabular}{lcccc}
\hline
 \textbf{Methods}& \textbf{ACC(\%)} & \textbf{P(\%)} & \textbf{R(\%)} & \textbf{F1(\%)} \\
\hline
MixHop-based & 83.68 & 87.60 & 78.47 & 82.78 \\
KD-based & 84.38 & 87.37 & 80.64 & 83.87 \\
\hline
\end{tabular}

\label{tab:comparison}
\end{table}

\section{Conclusion}
\label{sec:conculsion}
In this paper, we propose a novel model, Vul-LMGNNs, which integrates sequence and graph embedding techniques to detect vulnerabilities in function-level source code. Our approach leverages the code property graph representation of the source code as the primary input. Specifically, we utilize a pre-trained Program Language (PL) model to extract local semantic features from the code, which are then embedded as nodes in the graph using sequence-based embeddings. Subsequently, we employ a GGNN equipped with convolutional layers to effectively fuse heterogeneous information within the graph. Finally, our model jointly learns and predicts vulnerabilities by combining the PL model with the GGNN. To validate the effectiveness of Vul-LMGNNs, we conducted extensive experiments on four real-world datasets, which demonstrated its superior performance. We systematically explored trade-off parameters, fine-tuning of the PL model, and variations of GNN architectures. Our findings further emphasize the positive contributions of each module to the overall model performance. As part of interesting future work, we intend to explore more effective fusion networks for learning code representations and facilitating multiclass detection.

\bibliographystyle{unsrt}

\bibliography{main}



\end{document}